\documentclass[preprint,12pt]{elsarticle}

\usepackage{amssymb}

\usepackage{placeins}

\usepackage{mathrsfs}

\usepackage{subfig}

\usepackage[linesnumbered,ruled,vlined]{algorithm2e}

\usepackage{algorithmic}

\usepackage[utf8]{inputenc}

\usepackage{graphicx}

\usepackage{wrapfig}

\usepackage{soul}
\usepackage{xcolor}
\usepackage{float}

\usepackage[margin=1.25in]{geometry}

\journal{International Journal of Multiphase Flow}

\begin{document}

\begin{frontmatter}



%opening
\title{The Eulerian Lagrangian Mixing-Oriented (ELMO) Model}

\author[label1]{David P. Schmidt\corref{cor1}} \ead{schmidt@acad.umass.edu}
\author[label1]{Majid Haghshenas} \ead{mhaghshenas@umass.edu}
\author[label1]{Peetak Mitra} \ead{pmitra@umass.edu}
\author[label2]{Chu Wang} \ead{chu.wang@convergecfd.com}
\author[label2]{P. Kelly Senecal} \ead{senecal@convergecfd.com}
\author[label3]{Fabien Tagliante} \ead{ftaglia@sandia.gov}
\author[label3]{Lyle M. Pickett} \ead{lmpicke@sandia.gov}

\address[label1]{Mechanical and Industrial Engineering Department,
                 University of Massachusetts Amherst, MA}
                 
\address[label2]{Convergent Science Inc., Madison, WI}
\address[label3]{Sandia National Laboratories, Livermore, CA}
\cortext[cor1]{Corresponding author}


\begin{abstract}

    Past Lagrangian/Eulerian modeling has served as a poor match for the mixing-limited physics present in many sprays.  Though these Lagrangian/Eulerian methods are popular for their low cost, they are ill-suited for the physics of the dense spray core and suffer from limited predictive power.  A new spray model, based on mixing-limited physics, has been constructed and implemented in a multi-dimensional CFD code.  The spray model assumes local thermal and inertial equilibrium, with air entrainment being limited by the conical nature of the spray.  The model experiences full two-way coupling of mass, momentum, species, and energy.  An advantage of this approach is the use of relatively few modeling constants.  The model is validated with three different sprays representing a range of conditions in diesel and gasoline engines.
\end{abstract}
\end{frontmatter}

\section*{Introduction}

Spray modeling in CFD has long been dominated by Lagrangian-Eulerian methods.  The stochastic nature of the Lagrangian-Eulerian model has  permitted the efficient sampling of a sub-set of the actual drops 
 \cite{schmidt2018} while avoiding the frustrations of numerical diffusion \cite{dukowicz1980}.  The basic building block of these simulations has been the parcel, a statistically-weighted representation of a single droplet.  
 
 The predicted behavior of the spray depended on the droplet breakup models, drag models, collision models, and turbulent dispersion models.  Partial consideration could be made for the presence of neighboring drops in dense sprays \cite{balachandar2019}, but the essential nature of the spray model was drop-oriented.  It was hoped that if the individual droplet phenomena were modeled correctly, then accurate predictions of spray behavior would follow.  Under many situations this \textit{modus operandi} was successful, and for this reason, Lagrangian-Eulerian modeling remains a mainstay of applied CFD.

Lagrangian-Eulerian modeling has not proved to be ultimately satisfactory, especially for the high ambient density environments found in diesel engines.  The decline of this droplet-oriented view was precipitated by two problems.  The first was that the predictive power of Lagrangian-Eulerian spray modeling was found to be unsatisfactory without the adjustment of numerous model parameters \cite{precise2011}.  In fact, if model parameters must be frequently adjusted and cannot be \textit{a priori} known, then the model is not truly predictive.  Second, the predictive power is further compromised by the challenges in achieving grid-convergent solutions.  The computational requirements of observing convergence, as measured by mesh requirements \cite{senecal2014} and parcel count \cite{schmidt2018}, mean that most Lagrangian spray calculations are grossly under-resolved. 

The second was the successful mixing-limited model of Siebers \cite{siebers1999}.  Siebers found that spray evolution correlated with global parameters such as spray angle and that under engine-relevant conditions, interfacial details played no observable role in spray evolution.  Apparently, the interfacial area was so great that the mixing of ambient gas into the spray was the limiting factor in interfacial momentum and mass exchange.  With this realization, the droplet-oriented Lagrangian-Eulerian approach seemed less appropriate.  Instead, spray evolution resembled turbulent mixing between the liquid and the ambient gas.  In the absence of interfacial details, which Siebers showed were not germane, the view of the spray was more of a continuum, such as would occur in the turbulent mixing of a high density gas jet. 

The natural framework for modeling the mixing of two continuous fluids would be Eulerian-Eulerian modeling.  Because of the very high Weber number and minimal impact of interfacial details, interface tracking and capturing methods have proven very expensive for modeling diesel sprays.  Instead, Eulerian paradigms that model the turbulent mixing of the two phases have emerged as pragmatic approaches to capturing the mixing-limited behavior of sprays \cite{xue2014} \cite{blokkeel2003} \cite{garcia2013}. 

These Eulerian models provide close coupling between the phases without incurring the expense of resolving interfacial features.  In fact, the interfacial modeling of many Eulerian models is an appendage, where the interfacial details do not feed back into the flow field evolution \cite{xue2014}.  This construction is predicated on the assumption of mixing-limited flow.  Thermal evolution models, such as Garcia-Oliver\cite{garcia2013}, make the locally homogeneous flow (LHF) assumption advocated by Faeth \cite{faeth1983} with great success.  

Other advantages of the Eulerian diffuse-interface treatments is a lesser tendency for mesh dependency than Lagrangian-Eulerian modeling \cite{xue2014}.  While Lagrangian-Eulerian spray models do converge, they are only conditionally convergent--the user must employ a surprisingly large number of computational parcels in order to observe convergence \cite{schmidt2018}.  The Eulerian models also account for the volume occupied by the spray, which is typically neglected in Lagrangian-Eulerian simulations \cite{pakseresht2018}.  

Ultimately, however, direct injection engine simulations require a cost-efficient spray model for in-cylinder computations.  While Eulerian-Eulerian models have had many impressive successes in predicting near-field spray behavior \cite{iyer2005}\cite{luret2010}\cite{blokkeel2003}\cite{lebas2005}\cite{desantes2016}, these constructions are quite expensive to use for in-cylinder calculations.  The multi-scale nature of engine simulations becomes a severe problem because one must mesh a volume as large as the cylinder with resolution much smaller than the injector orifices \cite{xue2014}.  Some simplification is required in order to embed a mixing-limited spray model into an internal combustion engine (ICE) simulation.  

\begin{figure}[ht]
 \centering
 \includegraphics[width=0.7\textwidth]{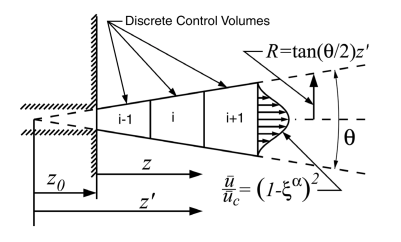}
 % MKcv.png: 401x242 px, 100dpi, 10.19x6.15 cm, bb=0 0 289 174
 \caption{The Eulerian control volumes used in the Musculus-Kattke model.  Taken from Musculus-Kattke(2009)}
 \label{fig:MKcv}
\end{figure}
% I am not sure I am recalling the details of these papers below correctly--must check

Simplified, one-dimensional, mixing-limited models have demonstrated substantial success with a greatly-reduced cost and complexity when compared to CFD.  A sketch of a one-dimensional Eulerian spray control volume is shown in Fig.\ \ref{fig:MKcv}.  Seminal examples include Naber and Siebers \cite{naber1996}, Musculus and Kattke \cite{musculus2009}, and Desantes et al. \cite{desantes2007, desantes20091d}.  These models all had several features in common.  First, they were essentially one-dimensional Eulerian calculations.  The work of Musculus et al.\ and Desantes et al.\ both included a simplified assumption of a radial profile within the spray.  The works of Musculus et al. and Desantes et al.  were both transient, allowing investigation of the start and end of injection.  The work of Desantes et al.\ considered vaporization and heat transfer, including details such as fugacity for calculation of liquid-vapor equilibrium. 

Ultimately, however, these models are not easily included into a multi-dimensional CFD computation.  There are at least three outstanding issues:
\begin{enumerate}
 \item How does one calculate the two-way coupling between the CFD solution and the spray evolution?  The reduced order models such as Musculus and Kattke assume a zero velocity ambient, but a CFD solution will produce a resolved gas velocity field that should be employed in the spray evolution.  A large part of this challenge is that the overlap of spray control volumes in a one-dimensional model with the underlying CFD mesh produces a complex interpolation problem.  Foremost of the difficulties becomes how to calculate the intersection between spray control volumes shaped like cone frustra and the underlying CFD mesh.  This intersection calculation would have to be repeated every time the spray control volumes or CFD mesh change.
 
 \item What if the spray bends?  This is usually less of a concern for diesel sprays, but if the model were applied to gasoline direct injection (GDI) sprays, the entrained momentum of the ambient gas would likely steer neighboring sprays together \cite{manin2015}.  However, if the spray is allowed to bend, then one does not \textit{a priori} know the location of the spray control volumes.
 
 \item In actuality, the spray angle is not a constant, as assumed by all the reduced order models above, but rather varies in time.  Spray angle tends to be higher at very low needle lifts \cite{moon2015}. In an Eulerian treatment of the spray model, this variation in spray angle would create control volumes that vary in time, greatly complicating the required numerical methods.

\end{enumerate}

Implementations of reduced-order Eulerian mixing-limited spray models into CFD codes have shown considerable promise \cite{abani2008}\cite{wan1999}\cite{versaevel2000}\cite{yue2017}\cite{yue2019}\cite{abraham1999virtual}.  However, these demonstrations have largely dodged the difficulties enumerated above.  In each case, their implementations assumed a constant spray angle and constrained the mass and momentum exchange to be mostly pre-determined.  As an area of improvement, we aspire to develop a fully-coupled model that is well-suited to time-varying spray angle.

A further goal is to investigate the effectiveness of a mixing-limited model beyond the traditional diesel engine conditions studied by Siebers \cite{siebers1999}.  This article will not only describe the implementation of the model, but test it under the much lower density and temperature environment found in gasoline direct injection (GDI) engines.  The actual region of applicability of the mixing-limited model is currently an open topic of study, hence there is no Weber number or Reynolds number map that can inform us when the model is applicable.  With empirical testing of the model, we may learn more about where it may be applied.

\section*{Features of the ELMO Model}
We propose a new model, the Eulerian-Lagrangian Mixing-Oriented (ELMO) Model.  This model represents similar physics to the work of Musculus and Kattke \cite{musculus2009} and Desantes et al. \cite{desantes2007} but is implemented in a CFD code with two-way coupling.  Thus, the predictions of the spray will adapt to the cylinder gas conditions in a more general way than prior implementations of mixing-limited models.

The foremost advance of the ELMO model is a predication on the mixing-limited physics.  The model is well suited for diesel sprays in the dense core, where Lagrangian-Eulerian models struggle.  The model may also show promise for gasoline sprays, at least where the mixing-limited hypothesis holds true.  Validation studies against diesel and gasoline measurements will indicate the applicability of this new model.  

The next advance of the ELMO model is the use of Lagrangian ``capsules'' that start out as liquid, but entrain ambient gas in order to become two-phase constructs.  The capsule is a moving, deforming control volume that follows a discrete mass of fuel.  Similar ideas were proposed by B{\'e}ard et al. (2000) \cite{beard2000} in their CLE model, with the idea of avoiding mesh dependency.  In a later development, full coupling has been included in the VSB2 spray model by K{\"o}sters and Karlsson \cite{kosters2016}.  Their model utilized two-phase Lagrangian entities with some similarities to the model proposed here.  The main difference between the VSB2 model and the present work is that the former emphasizes droplet size and a finite relaxation time to thermodynamic equilibrium.   In contrast, the present work is mixing limited, without reliance on interfacial details.  Rather than a finite relaxation time, the present model assumes instantaneous equilibrium, as in Pastor et al. \cite{pastor2012}.

Here, the ELMO model is, at its heart, a translation of the Musculus-Kattke model to a Lagrangian reference frame with an addition of equilibrium evaporative physics from Desantes et al.  The ELMO model is built on assumptions of inertial and thermal equilibrium, limited only by the entrainment rate of surrounding gas.

Another important distinction between ELMO and the past efforts that employed a two-phase control volume, such as the work of B{\'e}ard et al. (2000) \cite{beard2000} and K{\"o}sters and Karlsson \cite{kosters2016}, is that the geometry of the ELMO model's control volumes are based on the spray angle.  The spray angle is the dominant parameter for controlling the entrainment rate and provides a natural means to specify the breadth of the control volume based on the physical insights of Siebers \cite{siebers1999}.

An ELMO capsule has the following attributes:
\begin{enumerate}
 \item A capsule is a cone frustum representing a moving quasi one-dimensional control volume.  It has a finite extent and overlaps several gas phase cells, with which it exchanges mass, momentum, and energy.
 
 \item The mass of fuel in a capsule remains constant.
 
 \item A capsule contains both gas and liquid.  The gas is a combination of entrained gasses and evaporated fuel.
 
 \item The capsule has a cone angle that was determined at the time of the capsule's birth.  As a capsule moves downstream and grows, the cone angle is retained.  The capsule's cone angle is not necessarily the same as the current injection cone angle, which may be a function of time.
 
 \item A capsule has a velocity that is calculated from conservation of momentum.  As slow-moving air is entrained, the velocity of the capsule diminishes.
 
 \item A capsule is in thermodynamic equilibrium, with an assumption of locally homogeneous flow.  The liquid and gas are at the same temperature.
 
 \item The axial velocity and liquid volume fraction in the capsule follow a radial profile that evolves in the near-nozzle region.  This profile, used by the Musculus-Kattke model, is illustrated in Fig.\ \ref{fig:MKcv}
 
\end{enumerate}

\begin{figure}
 \centering
 \includegraphics[width=0.7\textwidth]{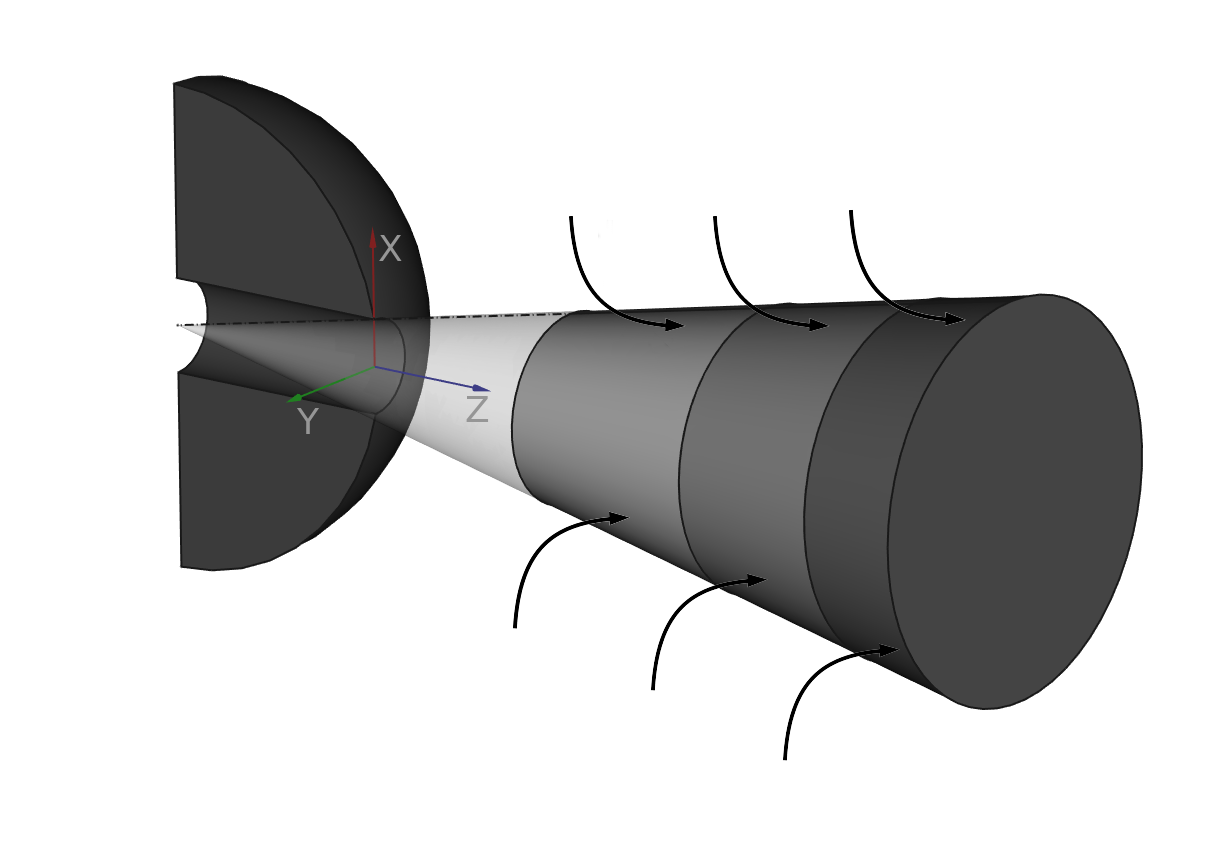}
 
 % ELMO.png: 1200x400 px, 72dpi, 42.33x14.11 cm, bb=0 0 1200 400
 \caption{A conceptual sketch of the ELMO model.  One capsule per time step is created at each injector orifice and approximately follows the formerly created capsule.  The arrows show gas entrainment into the capsules as they move from left to right. The capsule axial length grows increasingly shorter as it advects away from the injector tip, and the shortening of the length is compensated by the elongated radial extent, which follows the dispersion angle at capsule birth.   }
 \label{fig:elmo}
\end{figure}

A sketch of the ELMO concept is shown in Fig.\ \ref{fig:elmo}.  The curved arrows show entrained cylinder gas, which brings new enthalpy for vaporization and mass, which tends to diminish the velocity of the capsule.  A new capsule is generated every time step, so that they emerge from the injector in series.  

The decision to use a Lagrangian reference frame has several benefits.  The first, is that one need not construct a spray mesh that overlays the gas phase mesh.  If the spray angle varies as a function of time or the spray bends, such a spray mesh would need to move and deform.  Instead, we dynamically calculate overlap between capsules and the underlying mesh by using stochastically distributed points within the capsule.  

A second benefit is the ability to re-use some of the computational infrastructure that was developed for Lagrangian-Eulerian spray modeling.  For example, the stochastically distributed points mentioned above are located in the gas phase mesh using the same software functionality developed for locating parcels.

Because the number of capsules is low compared to the number of parcels used in a typical Lagrangian-Eulerian spray calculation, the ELMO model imposes a very small computational burden.  There is one capsule generated per orifice per time step.  Each existing capsule is updated for the time step without sub-stepping, until the capsule contains so little liquid that it transitions to a very sparse assortment of Lagrangian parcels.   

Most importantly, the ELMO model has few arbitrary physical constants, providing little temptation to tune the model.  This makes the predictions reproducible and predictive.  The ELMO model is equally applicable to RANS models and LES closures.

The next sections describe the governing equations and the method of implementing the ELMO model.  The ELMO model was incorporated into the CONVERGE 2.4 CFD code for application to spray simulations.  The details of the CONVERGE code are described elsewhere \cite{senecal2014} \cite{converge_manual} and so the emphasis of the following text is on the spray modeling.

\section*{Governing Equations}
The governing equations are derived using a control volume analysis.  Each capsule is a cone frustum, with a front side that extends a distance $z'_{fs}$ from the virtual origin and a distance of $z'_{bs}$ from the virtual origin (see Fig.\ \ref{fig:MKcv}).  The control volume encompasses the entire capsule: liquid fuel, vapor fuel, and non-condensible gas.  The total mass and volume of the capsule represent the sum of the three components.  The liquid, vapor, and non-condensible gas are represented by the subscripts $l$, $v$, and $a$, respectively.

\begin{equation}
 \label{eq:mass}
 m = m_l + m_v + m_a
\end{equation}

\begin{equation}
 \label{eq:volume}
 V = V_l + V_v + V_a
\end{equation}

The inclusion of gaseous species in the control volume means that the capsule control volume overlaps with the Eulerian CFD control volumes.  A separate mathematical analysis is thus required for calculating the phase interactions.  
Underlying all of these equations are assumptions of equilibrium.  Specifically, the phases all move at the same velocity and experience equal pressure.  All phases are in thermal equilibrium as well.  

The mass of a capsule continuously increases due to the entrainment of air.  Each timestep, a capsule's mass, $m$, increases by the entrained mass.

\begin{equation}
\label{eq:MassConsv}
 m^{n+1} = m^{n} + \Delta m
\end{equation}

This entrained gas, may include some momentum, $M$, due to the velocity of the entrained gas, $u_{gas}$. 

\begin{equation}
 \label{eq:MomConsv}
 M^{n+1} = M^{n} + u_{gas} \Delta m
\end{equation}

In the present implementation, this entrained momentum is neglected, as in the work of Musculus and Kattke \cite{musculus2009}.  This assumption results in a constant capsule momentum.  Like the implementation of Musculus and Kattke, the consideration of a radial velocity and density profile is included here, which has implications for momentum.  As shown in Eqn.\ \ref{eq:ubar}, the correlation of the distribution of mass and momentum results in a parameter $\beta$ in the expression for momentum.  This parameter, calculated from the root of a fourth-order polynomial following Musculus-Kattke \cite{musculus2009}, arises from the fact that the velocity is higher near the spray axis, where most of the mass is concentrated. 

\begin{equation}
 \label{eq:ubar}
 M = \beta m \bar{u}
\end{equation}

As the capsule advances each time step, the center of mass $z$ is updated according to a simple explicit Euler time integration.  Again, the correlation parameter $\beta$ appears due to the correlation between mass and velocity distribution.

\begin{equation}
    z^{t+\Delta t} = z^t + \beta \bar{u} \Delta t
\end{equation}

The capsule changes shape as it evolves, with an increasing radial extent and a diminishing axial extent.  The reduction in axial extent is due to the reduction in velocity with increasing distance from the injector orifice.  This same reduction in velocity represents a deceleration of the center of mass of each capsule.  The derivation of the front side ($fs$) and back side ($bs$) location of each capsule begins with the assumptions that these capsule boundaries lie halfway between the centroid $z$ of a capsule at its current position and next position for the front side or halfway between the current centroid location and previous location for the back side.  Expressing this mathematically:
\begin{equation}
    {fs}^t = mean\left( {z}^t , {z}_{est}^{t+\Delta t} \right)
\end{equation}

\begin{equation}
    {bs}^t = mean\left( {z}^t , {z}_{est}^{t-\Delta t} \right)
\end{equation}

Here, the subscript ${est}$ indicates an estimate described below.  Because of the expected $1/x$ decay of velocity with distance from the injector, a harmonic mean is employed in the model instead of an arithmatic mean. These constraints provide a nearly contiguous arrangement of capsules, as illustrated conceptually in Fig.\ \ref{fig:elmo}. 

The extent of the capsule is determined using the concept of a fictitious train of identical capsules.  The centroid positions at the previous and next capsules are estimated using a first-term Taylor series expansion, for example, in Eqns.\ \ref{eq:TaylorSeries} and \ref{eq:TaylorSeries2}.  Note that $\Delta t^0$ is not the current time step, but rather the time step at the capsule's birth because these estimates of a capsule's extent are based on the size of the capsule, which is a quantity that evolves continuously from its creation and is independent of the current time step.

\begin{equation}
    {z}_{est}^{t+\Delta t} = z^t + \Delta t^0 \frac{\partial z}{\partial t} = z^t +  \Delta t^0 \beta \bar{u} 
    \label{eq:TaylorSeries}
\end{equation}

\begin{equation}
    {z}_{est}^{t-\Delta t} = z^t - \Delta t^0 \frac{\partial z}{\partial t} = z^t - \Delta t^0 \beta \bar{u} 
    \label{eq:TaylorSeries2}
\end{equation}

Once the location of the front and back sides are updated, the geometry of the capsule is fully determined and the kinematics of the advection process is complete.  Next, the thermodynamics are calculated. Here, we largely follow the thermodynamic analysis of Desantes et al. \cite{desantes2007} who calculated the vaporization of single component fuels under mixing-limited assumptions.  In the present case, there are some additional constraints that require iterative solution of the equations below.

The vaporization is driven by the rate of entrainment of ambient gas at a temperature of $T_{\infty}$ and a pressure of $p_{\infty}$.  All ambient quantities seen by the capsules are evaluated at these conditions.  All gas species are treated as ideal gasses for the sake of simplicity, though we acknowledge that these assumptions are not defensible at high pressures.  As will be shown below, we can reproduce the results of Desantes et al, suggesting that the consequent error may be acceptable. 

First, we calculate the vapor pressure, including the Poynting correction for an ideal gas at high pressure \cite{desantes2007}.  In the equations below, the mass fraction of fuel, $Y_f$ is defined as the mass of fuel divided by the total mass of the capsule.  The mass fraction of fuel vapor, $Y_{fv}$, is defined as the mass of fuel vapor divided by the capsule mass.  The uncorrected vapor pressure is denoted as $p^0_{vap}$.

\begin{equation}
\label{eq:Poynting}
 p_{vap} = p^0_{vap} * exp \left[ v_{liq} \frac{p_\infty - p^0_{vap} }{R_{gas} T_\infty} \right]
\end{equation}

The partial pressures for a mixture of ideal gasses is proportional to mole fractions, allowing the derivation of Eqn. \ref{eq:Yfv}.  Here $MW_{fuel}$ is the molecular weight of the fuel species and $MW_\infty$ is the molecular weight of the entrained gas.

\begin{equation}
 \label{eq:Yfv}
 Y_{fv} = \left( 1 - Y_f \right) \frac{MW_{fuel}}{MW_\infty} 
 \frac{1}{\frac{p_\infty}{p_{vap}} - 1}
\end{equation}

By assuming that the evolution of species concentration and enthalpy due to entrainment evolves in an analogous fashion, per the unity Lewis number assumption, Desantes et al. derived the following relationship between fuel vapor fraction and fuel fraction.  Here, $h_a$ represents the enthalpy of the ambient gas, either at $T_{\infty}$ or the temperature of the capsule, $T$.  The denominator of the right side, $h_{fg}$ is the enthalpy of vaporization. 

\begin{equation}
\label{eq:analogy}
 \frac{Y_{fv}}{1-Y_f} = \frac{h_{a}(T_{\infty}) - h_a(T)- \frac{Y_f}{1-Y_f}\left(h_f(T) - h_f(T_{inj}) \right) }
 {h_{fg}}
\end{equation}

Though the local gas conditions are known, the far-field conditions, such as $T_{\infty}$, are not readily available in a CFD context.  As described in a later section, surface Monte-Carlo points are used for estimation of the surrounding conditions by sampling thermodynamic variables at the periphery of the capsule frustrum.

% , but some effect of the proximity of the spray is still evident.  For example, if the surface of the capsule were in the presence of undisturbed gas, the mass fraction of evaporated fuel would presumably be zero, when in fact, the value observed in the gas phase at the outer surface of the capsules can exceed 0.2.  Hence, a correction is made to the value of $h_a(T_{\infty})$, as indicated by Eqn. \ref{eq:hcorr}.  The correction presumes the one-to-one correspondence between enthalpy and fuel mixture fraction established by Desantes et al. \cite{desantes2007}, The suffix $surf$ denotes the values observed at the outer surface of the capsule.  An illustration of the correction is shown in Fig.\ \ref{fig:hcorr}.

% \begin{equation}
%   h_{a}(T_{\infty}) = \frac{h_{a}(T_{surf}) - h_f\left(T_{inj} \right) Y_{fv_\infty} } {1.0-Y_{fv-surf}}
% \label{eq:hcorr}
%   \end{equation}

% \begin{figure}[ht]
%  \centering
%  \includegraphics[width=0.8\textwidth]{correctionToAmbientEnthalpy.png}
%  % frustumWithEntrainment.png: 816x1056 px, 96dpi, 21.59x27.94 cm, bb=0 0 612 792
%  \caption{An illustration of how ELMO extrapolates to find the far-field temperature using the observed conditions at the periphery of the capsule (point).}
%  \label{fig:hcorr}
% \end{figure}

With the inclusion of Eqn. \ref{eq:analogy} the equilibrium state of the capsule is fully determined.  However, the goal of the ELMO model is not just to predict the evolution of the spray, but to provide the source terms required for the gas phase Navier-Stokes solver.  The intellectual challenge presented by the ELMO model is that the capsule is a two-phase entity.  Hence, the gas portion of the capsule control volume overlaps with the control volumes of the gas phase solver--the gaseous species are represented in both.  

In order to construct disjoint control volumes, to avoid this conceptual difficulty, we consider only the liquid phase portion of the capsule.  Then, the gain of the gas phase is simply the loss of the liquid phase.  For example, consider the mass source due to vaporization, where the evaporated vapor mass during one time step is represented $\Delta m_{evap}$.  Here, $\Delta m_{evap}$ is positive for mass leaving the liquid.

\begin{equation}
\label{eq:massLiq}
 S_m = - \left( m_l^{n+1} - m_l^n \right) = \Delta m_{evap} 
\end{equation}

The source term gained by the gas phase conservation of mass equation is simply $\Delta m_{evap}$.  Similarly, the momentum gained by the gas phase is the momentum lost by the liquid phase.

\begin{equation}
 \label{eq:momLiq}
 S_M = -\left( M_l^{n+1} - M_l^{n} \right) = m_l^n \beta ^n \bar{u}^n - m_l^{n+1} \beta^{n+1} \bar{u}^{n+1}   
\end{equation}

Note that the source term in Eqn.\ \ref{eq:momLiq} represents only the axial component of the momentum source.  Some contribution of radial momentum is also appropriate for locations off of the spray axis.  The model constructs a unit vector that points from the virtual origin to the location where the source term is being deposited.  A radial component is computed such that the momentum source vector is co-linear with this unit vector.

Like mass and momentum, the energy lost by the continuous phase is equal to the energy gain from the liquid phase.  However, because CONVERGE solves an energy equation that only tracks the sensible portion of energy, the actual implementation is given by Eqn. \ref{eq:enthSource}.  The first term on the right side of the equation represents sensible heat transfer absorbed from the surroundings by temperature change, the second term the enthalpy of vaporization, and the third counts for the energy automatically gained by the addition of vapor mass at the continuous phase temperature.

\begin{equation}
    \label{eq:enthSource}
    S_e = m_l^n c_{p,l} \left(T^n - T^{n+1} \right)
    -\Delta m_{evap} h_{fl}
    -\Delta m_{evap} e_g
\end{equation}

\section*{Algorithm}

An explicit Euler time-marching approach is employed for the kinematic variables using the same time step as the overall CFD solver. Equations \ref{eq:MassConsv} through Eqn.\ \ref{eq:TaylorSeries} are updated and the new front side and back side positions are calculated. These new positions determine the volume of each capsule.  The new mass, however, is not known until the entrained gas mass is calculated.  Some of the increase in volume will be filled with non-condensible gas and some will be filled by newly evaporated fuel vapor.

The calculation of the thermodynamic state, however, poses greater difficulties, due to the coupling of the state variables and the non-linearity of the equations.  Equations \ref{eq:Poynting} through Eqn.\ \ref{eq:analogy} are solved iteratively, while maintaining the constraint of Eqn.\ \ref{eq:volume} according to Amagat's law. 

\begin{figure}
    \centering
    \includegraphics[width=0.4\textwidth]{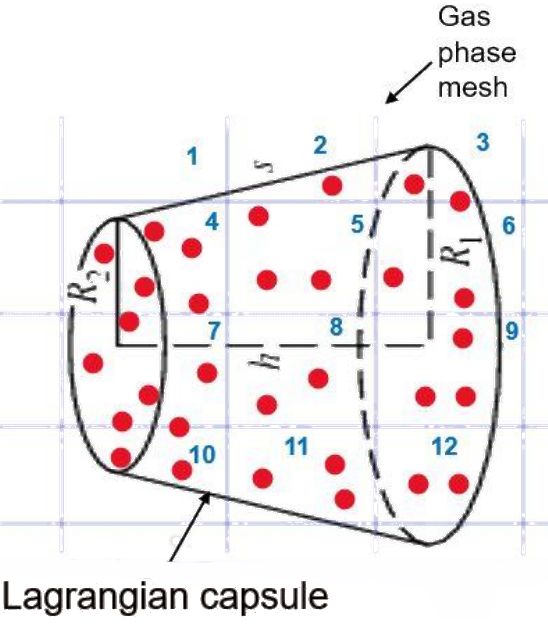}
    \caption{ELMO Capsule overlap with Eulerian mesh and associated Monte-Carlo points.  The actual implementation uses a thousand points distributed through the capsule volume.  Here, only 31 points are shown for clarity.  The blue numerals represent cell identification numbers and the black dimensions show the cone frustra extent.}
    \label{fig:capsule_MC}
\end{figure}

The coupling terms identified in Eqns.\ \ref{eq:massLiq} through \ref{eq:enthSource} must be distributed to the appropriate gas phase cells.  Depending on the mesh resolution and the size of the capsule, each capsule volume may overlap numerous gas phase cell volumes.  The intersection of these volumes are, in general, complex non-convex three-dimensional shapes.  Though geometric intersection algorithms do exist \cite{menon2011}, they are complex and time consuming.  Instead, a Monte-Carlo integration is used to distribute the mass, momentum, and energy to each gas phase cell.  Monte-Carlo integration provides a probabilistic integration of the source terms of the gas phase equations and is also used for interpolating from the gas phase to provide average gas properties in the capsule evolution equations.  This integration is illustrated in Fig.\ \ref{fig:capsule_MC}.

This distribution process has two major steps, the creation of Monte-Carlo points and the assignment of weights for source terms.  A third step, the transition from capsules to a more traditional Lagrangian-Eulerian framework is often employed near the end of a capsule's useful life.

 \subsection*{Creation of Monte Carlo points}
 A large number of points are created within each capsule using an acceptance/rejection algorithm. The probability distribution within space is non-uniform, with greater weighting towards the center, per the assumption of a radial profile of mass and velocity distribution.  Because of the mathematical challenge of distributing points at random within a conical frustum, we distribute the points within a cylinder that encloses the frustum.  The diameter of this cylinder is equal to the diameter of the front side of the capsule.  Any points that fall outside of the frustum are rejected and discarded. 
 
 \subsection*{Calculation of Weights}
 Each point is located in the mesh using the same algorithm used to find the cell in which a Lagrangian parcel might reside.  Then, the algorithm loops over all of the $N$ points that were accepted in the above step.  For each of these Monte Carlo points, a fraction of $1/N$ of the source term is deposited into the cell which contains the point.

This algorithm will, in a stochastic sense, distribute the source terms according to the volume fraction of the capsule represented by a gas phase cell.  For example, if a gas phase cell overlaps with nine percent of the capsule's volume, then approximately nine percent of the Monte Carlo points are expected to land within that gas phase cell.  The cell will receive, on average, nine percent of each source term.  For example, in Fig.\ \ref{fig:capsule_MC} three of the thirty-one red points fall in cell 11, indicating that a fraction equaling 3/31 of any source should be distributed to this cell. This stochastic approach obviates the need for calculating intersections.

These interior points are used only for distributing source terms. A separate set of points that are scattered on the surface of the capsule are employed for calculating the conditions of ambient gas.  These surface points are used to produce a weighted average of the ambient temperature and fuel vapor mass fraction.  

\subsection*{Transition}
At some point, the capsule approach is no longer needed.  The spray becomes sufficiently dilute, such that treating the spray as a collection of individual droplets is fully sufficient.  At this point, it becomes convenient to transition the capsules to traditional Lagrangian droplets.  Though the transition may not always be necessary, it has several advantages.  First, if the spray might impinge on a surface, the ELMO concept would not have a natural connection to existing spray impingement models.  Converting the spray to Lagrangian parcels enables the use of existing functionality to capture parcel impact on walls.  

Second, the assumption of the spray angle remaining constant far downstream of the point of injection is questionable.  At some point, the spray will deviate from a simple conical shape.  The transition relaxes this assumption and allows natural aerodynamic drag to control the shape of the spray tip.  

Third, the current implementation does not allow the velocity direction of the capsule to deviate from the injector hole axis.  However, in Lagrangian particle tracking there is no such limitation.  Consequently, the transition to Lagrangian points allows the model to capture spray bending, which is expected in some multi-hole injectors and in the presence of strong cross-flow.  

The conversion from capsule to Lagrangian parcels occur when any of the following criteria are met:  
\begin{enumerate}

    \item The liquid volume fraction of the capsule is below 0.5 percent
    \item More than 95 percent of the fuel is in vapor form
    \item The capsule velocity has diminished below 10 $[m/s]$

\end{enumerate}

These thresholds are admittedly arbitrary.  The first is based on the fact that once the liquid volume fraction is very low, the spray is well-represented by droplets evolving in relative isolation, making the typical assumptions of the Lagrangian models apropos.  By assuming a single drop size and a uniform lattice arrangement of droplets in three-dimensional space, one can translate liquid volume fraction (LVF) into inter-drop spacing. At an LVF of 0.005, the center-to-center distance between droplets is approximately six diameters.  Past direct numerical simulations of neighboring droplet interactions indicate that at such distances, the droplets evolve in a largely independent fashion \cite{quan2011}.  Hence, at a LVF of less than 0.5 percent, Lagrangian particle tracking is expected to perform satisfactorily. 

The second conversion criterion is based on the idea that if 95\% of the mass has evaporated, the ELMO model's responsibilities are largely complete.  The last transition criterion, based on velocity, is intended to avoid applying the model at such low velocities that the mixing-limited hypothesis would be absurd.  In the test cases found later in this paper, the LVF criterion was usually the operative factor.  The velocity criterion did not come into play.

At the point of transition, the liquid mass is divided evenly among an arbitrarily large number of parcels who are located at random points within the capsule.  The velocity is randomly selected from within the cone angle of the capsule with a magnitude equal to that of the capsule.  

The initial droplet size of the parcels presents a challenge, since droplet size has no role in mixing-limited models and does not exist within the capsule concept.  Instead, we hypothesize that the droplets are the largest expected radius based on a critical Weber number of 6, as observed by Pilch and Erdman for a wide range of Ohnesorge number \cite{pilch1987}.  Consistent with the locally homogeneous flow assumption, the gas and liquid are expected to be moving at roughly the same velocity, such that the velocity scale in the Weber number calculation is based instead on the expected turbulent fluctuations.  This value is calculated based on the observation by Kobashi et al. that, in numerous sprays and gas jets, both in the literature and their own experiments, the turbulence intensity is roughly 0.2 \cite{kobashi2018} near the centerline. The 20\% turbulence intensity is a robust feature, persistent over a wide range of downstream pressures and injection pressures.  While this estimate of droplet size is perhaps speculative, Lagrangian particle tracking is only invoked once the spray has transferred the large majority of its mass, momentum, and energy to the gas.

Once the transition to Lagrangian parcels is complete, primary atomization is presumed complete, and the parcels evolve with typical spray models appropriate for secondary atomization of dilute sprays. For all of the tests in the current work, the secondary spray models were the TAB breakup model \cite{orourke1987}, the No Time Counter (NTC) droplet collision model \cite{schmidt2000}, and the Frossling evaporation model \cite{frossling1938}. 

\section*{Results}

\begin{figure}[h!]
 \centering
 \includegraphics[width=0.65\textwidth]{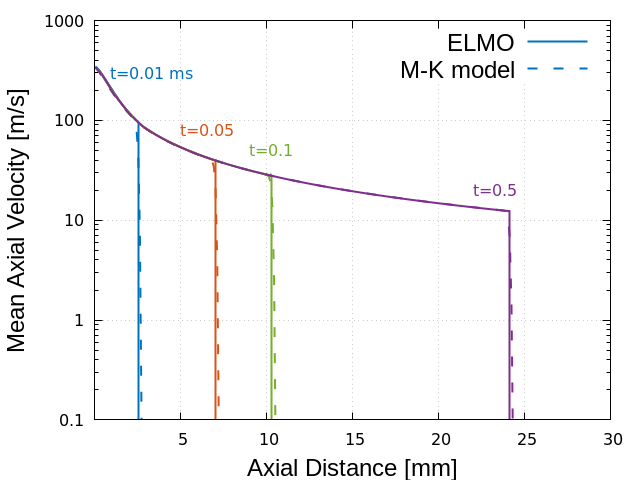}
 \caption{ Fuel jet penetration compared to Musculus-Kattke model \cite{musculus2009}.}
 \label{fig:MK_Ver}
\end{figure}

The results presented here comprise two categories:  verification and validation.  Verification provides evidence that the equations are being solved correctly and validation provides evidence that the model represents actual sprays.  Verification results are presented first, beginning with comparison to the Musculus-Kattke model.  This comparison is intended to show that the inertial portion of the ELMO model is being solved correctly.  The test case is based on the Engine Combustion Network (ECN) Spray A test case \cite{ECNweb}.  The conditions are given in Table \ref{table:sprayAexp}.  Note that the Musculus-Kattke model does not account for vaporization, but is only sensitive to ambient gas density.  The ELMO model, in contrast, includes vaporization effects.  So the temperature of the Spray A condition is divided by three and the pressure is multiplied by a factor of three in order to maintain the correct gas density while avoiding the issue of vaporization. This allows testing of the model without changing the code. The verification is shown in Fig.\ \ref{fig:MK_Ver}.

\begin{figure}[h]
 \centering
 \includegraphics[width=0.65\textwidth]{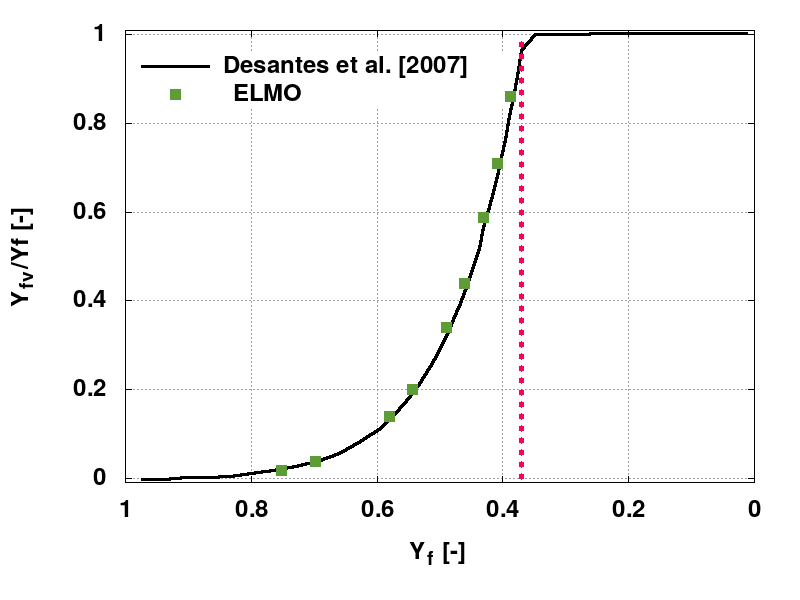}
 % ComparisonToDesantesFig5.png: 792x612 px, 72dpi, 27.94x21.59 cm, bb=0 0 792 612
 \caption{With a static, undisturbed flowfield, the current implementation reproduces the results of Desantes et al.  The abscissa represents the mass fraction of fuel and the ordinate is the fraction of fuel in the vapor phase.}
 \label{fig:DesantesComparison}
\end{figure}

The thermodynamic implementation was tested as a verification measure to ensure that it was faithful to the thermodynamics of Desantes et al. \cite{desantes2007}.  In their zero-dimensional model, there is no moving gas and no spatial variation of the surroundings, so the coupling of the ELMO model was temporarily disabled in order to reproduce the conditions of the  Desantes et al. calculations.  To make a fair comparison with the work of Desantes et al., the ELMO code must see a uniform ambient condition undisturbed by the spray.  The results are shown in Fig.\ \ref{fig:DesantesComparison}. As a capsule advects and grows, it entrains the surrounding gas.  This entrainment reduces the mass fraction of fuel from unity, at injection.  As the entrainment proceeds, the mass fraction of fuel in vapor form increases, until at a value slightly less than 0.4, for these particular conditions, the fuel is completely vaporized.  

\begin{table}[h]
 \centering

\begin{tabular}{|l|l| }
\hline
Numerical setup & \\
\hline
CFD code & CONVERGE V2.4 \\
Type grid & AMR and fixed embedding \\
Base grid & 2 mm \\
Embedding level for AMR and fixed embedding & 3 (on velocity gradient) \\
Maximum grid resolution & 0.25 mm \\
\hline
Models & \\
\hline
Turbulence & RNG $k-\epsilon$ \\ 
Breakup model & Taylor Analogy Breakup (TAB) \cite{orourke1987}\\
Droplet collision & No Time Counter (NTC) \cite{schmidt2000}\\
Vaporization & Frössling \cite{frossling1938} \\
\hline
\end{tabular}
\caption{Common numerical setup features for all three test cases.  Note that the droplet models are only invoked after transition.}
 \label{table:num_setup}

\end{table}

For validation, three Engine Combustion Network (ECN) cases were considered: Spray A, Spray H, and Spray G \cite{kastengren2012} \cite{duke2017} \cite{kosters2016}.  Because of the mixing-limited nature of the model, the validation did not include consideration of drop sizes, but instead focused on global spray quantities, such as vapor penetration, radial fuel distribution, and gas motion.  Drop size is not supposed to be a significant spray characteristic in mixing-limited sprays. The ELMO model was employed, without any adjustment to model parameters, in all validation cases.  The common features of the simulations are shown in Table \ref{table:num_setup}.

\begin{table}[h]
 \centering

\begin{tabular}{|l|l| }
\hline
Parameter &Value \\
\hline
Fuel & dodecane \\
Diameter & 90 [$\mu$m] \\
Duration of Injection & 1.54 [ms] \\
Mass Injected & 3.46 [mg] \\
Ambient Pressure & 6 [MPa] \\
Ambient Temperature & 900 [K] \\
Plume Cone Angle & 21.5 [degrees] \\
Fuel Temperature & 373 [K] \\
Coefficient of Area & 0.98 \\
Initial turbulent kinetic energy & $5.02 \cdot 10^{-4} m^2/s^2$ \\
Initial turbulent dissipation & 0.0187 $m^2/s^3$\\
\hline
\end{tabular}
\caption{Spray A experimental parameters \cite{pickett2011}}
 \label{table:sprayAexp}

\end{table}

The first case, Spray A, is a single axial hole injector with a rounded inlet.  This is a highly evaporating non-combusting case where the conditions are representative of diesel fuel injection.  The injection parameters are summarized in Table \ref{table:sprayAexp}.  For this case, a variable rate of injection profile was employed.  The three-dimensional simulation used a mesh with an initially uniform cell size of 2 $mm$ that is refined using fixed embedding around the spray. A cylindrical area is set by the nozzle with a radius of 2 $mm$ and length of 7 $mm$ for embedded refinement, while the embedding scale is set to 3.  Additionally, the grid is automatically refined by the CONVERGE code based on the magnitude of gas velocity. The initial turbulent kinetic energy and dissipation are estimated from the experimental data \cite{pickett2011}, combined with an assumption that the turbulence length scale is equal to the nozzle diameter.   

\begin{figure}[h]
    \centering
    \includegraphics[width=13 cm]{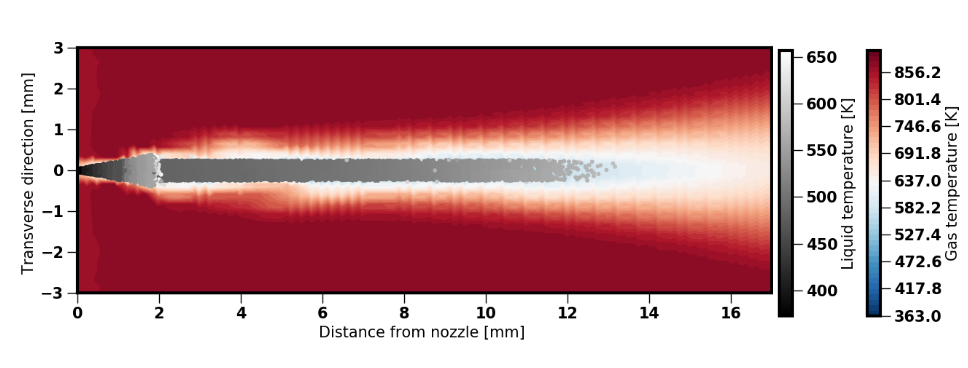}
    \caption{Spray A at 1 $ms$ after the start of injection.  The capsules are rendered as a sub-sample of the Monte-Carlo points.  Parcels and capsules are rendered on the same temperature scale.}
    \label{fig:sprayAsnap}
\end{figure}

A snapshot at 1 $ms$ after the start of injection is shown in Fig.\ \ref{fig:sprayAsnap}.  For visualizing sprays simulated with ELMO, a sub-sample of the Monte-Carlo points are selected for post-processing.  The figure shows a discontinuity in the spray angle at the transition point.  Since capsules are cone frustra, simulating the entire spray with ELMO would force a strictly conical shape on the spray, whereas transition provides a more natural treatment of the tip region.  The location of transition varies, but by the end of injection, the distance is 1.9 $mm$ from the injector orifice.  The disadvantage, however, of the transition process is that the drag on the liquid changes abruptly from the stipulated drag of the Musculus-Kattke model to the drag calculated by the CFD solver, resulting in the discontinuous shape.  Similarly, the governing equations for liquid temperature abruptly shift at the point of transition producing a rapid change in liquid temperature.

\begin{figure}[H]
    \centering
    \subfloat[]{{\includegraphics[width=9.5cm]{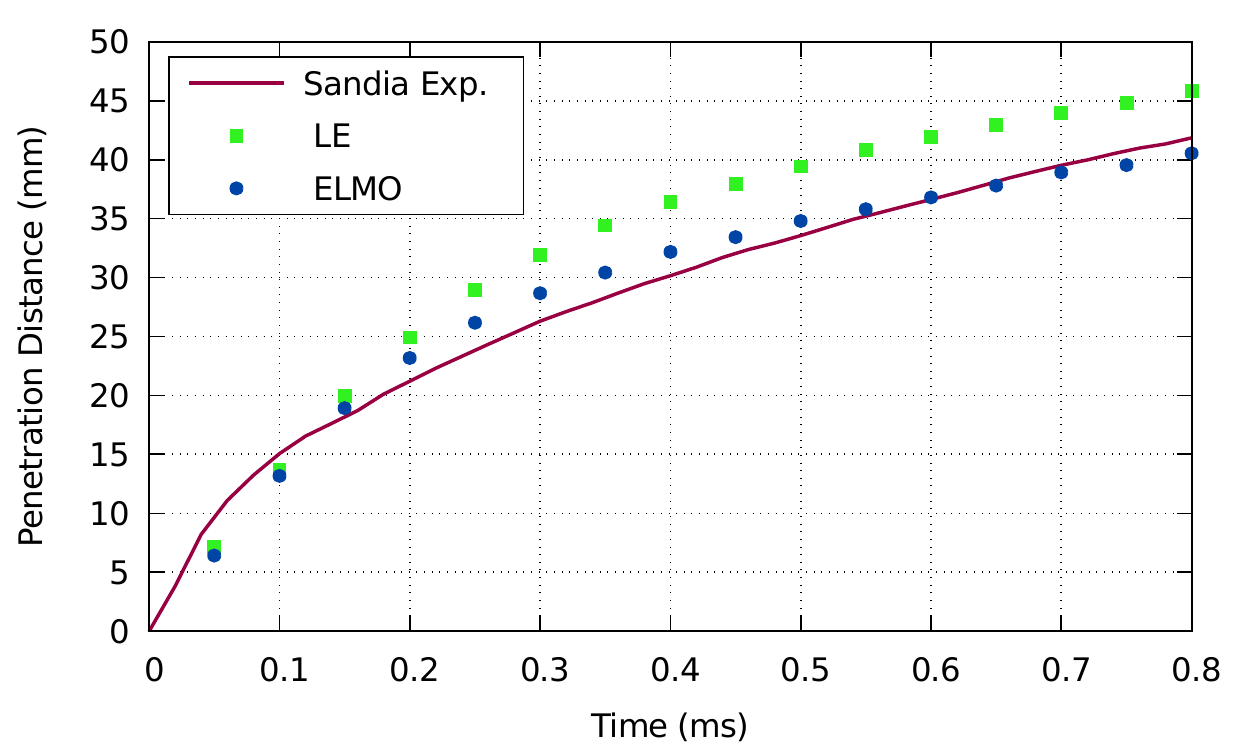} }}%
    \qquad
    \subfloat[]{{\includegraphics[width=10cm]{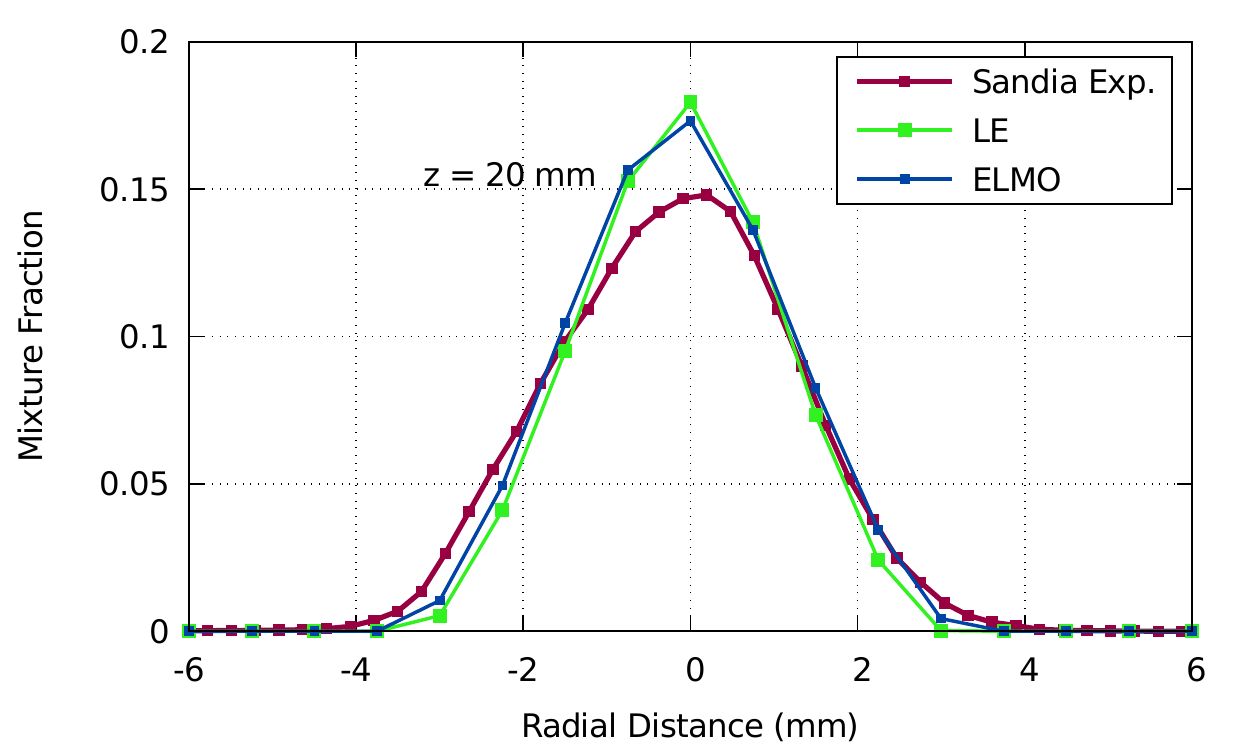} }}%
    \qquad
    \subfloat[]{{\includegraphics[width=9.5cm]{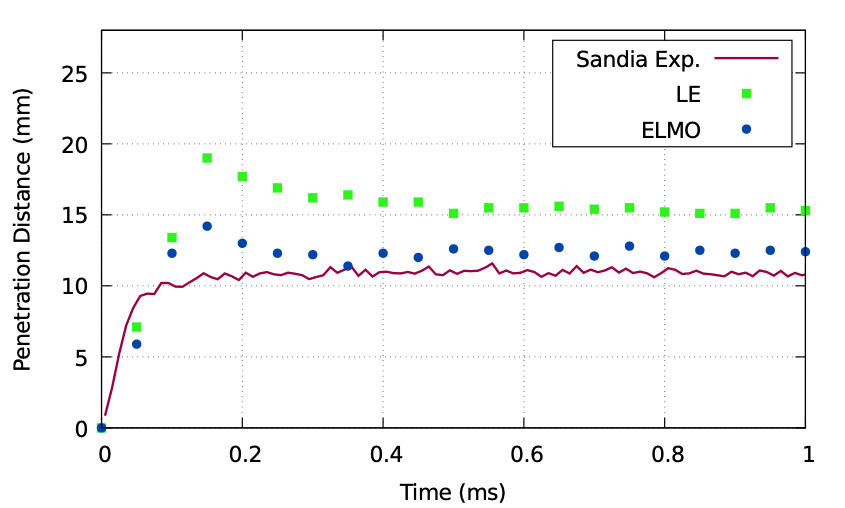} }}%
    \caption{Spray A simulation compared against experimental result of Pickett et al \cite{pickett2011}. a) vapor penetration, b) transverse mixture fraction distribution c) liquid length.  Because the experiment has imperfect symmetry, the experimental data for transverse mixing fraction have been shifted horizontally to align with the simulation.  }%
    \label{fig:SpA}%
\end{figure}

For validation, the predictions of the ELMO model were compared to experimental data from Pickett et al \cite{pickett2011}.  Fig.\ \ref{fig:SpA} shows the spray vapor penetration versus time (a), the transverse distribution of mass at 20 $mm$ downstream of injection at the end of injection (b), and spray liquid length (c). Without model adjustment, the penetration is predicted by the ELMO model.   

For comparison, a Lagrangian-Eulerian simulation result is also included in Fig. \ \ref{fig:SpA}.  The model inputs were specified as a part of an example case that comes with CONVERGE 2.4.  This case employs a dynamic droplet drag model \cite{liu1993}, the Fr{\"o}ssling evaporation model \cite{frossling1938}, the KH-RT atomization model \cite{beale1999}, and the NTC droplet collision model \cite{hou2006}.  These models were used, without adjustment of parameters, for this case and for subsequent cases shown later in this paper.

The transverse profiles shown in Fig.\ \ref{fig:SpA} (b) permit comparison of the fuel dispersion in the radial direction.  Comparison to the liquid length indicates that the fuel is entirely vaporized well before these measurement locations.  Due to slight asymmetries in the fabrication of the Spray A nozzle, the experimental profiles are not perfectly symmetrical nor are they centered on a radial distance of zero.  Small amounts of asymmetry occur in the ELMO results due to the stochastic nature of the spray models.  At 20 $mm$ ELMO predicts the bell-shaped distribution with the approximate width and height of the experimental measurements.  Fig.\ \ref{fig:SpA} (c) shows the spray liquid length predicted by LE, ELMO, and experiments of Sandia national lab. The liquid length has been computed based on projected liquid volume fraction using a $0.2\cdot 10^{-3} mm^3$ liquid volume per $mm^2$ area threshold, like the experiments.  ELMO shows a good prediction compared to LE and experimental data. 

The next test case is the ECN Spray H injection measurements \cite{idicheria2007}.  The Spray H injector is a single axial hole injector with a sharp entrance, more likely to cavitate than Spray A.  Indicative of this sharper nozzle entrance, the value of the coefficient of area is lower than Spray A.   The injected fuel is n-heptane, which is more volatile than dodecane, making Spray H a substantially different target than Spray A.  The conditions used in the simulation are given in Table \ref{table:sprayHexp}.  The rate of injection was described by an idealized profile, rather than assumed constant. Like the previous simulation, a uniform 2 $mm$ mesh initial resolution was used, and then adaptive mesh resolution took control as the spray developed. 

The most critical single quantity amongst the parameters in Table \ref{table:sprayHexp} is the spray angle.  The value of 23 degrees was recommended by Pickett et al. \cite{pickett2011} as appropriate for modeling, though their high-sensitivity experiments indicated an angle of 24 degrees.  The latter value was chosen for the present simulations.

\begin{table}[h]
 \centering

\begin{tabular}{|l|l| }
\hline
Parameter &Value \\
\hline
Fuel & n-heptane \\
Diameter & 100 [$\mu$m] \\
Duration of Injection & 6.8 [ms] \\
Mass Injected & 17.8 [mg] \\
Ambient Pressure & 4.33 [MPa] \\
Ambient Temperature & 1000 [K] \\
Plume Cone Angle & 24 [degrees] \\
Fuel Temperature & 373 [K] \\
Coefficient of Area & 0.86 \\
Initial turbulent kinetic energy & $5.02 \cdot 10^{-4} m^2/s^2$ \\
Initial turbulent dissipation & 0.0176 $m^2/s^3$\\
\hline
\end{tabular}
\caption{Spray H Experimental Parameters \cite{pickett2011}}
 \label{table:sprayHexp}

\end{table}

\begin{figure}[H]
    \centering
    \subfloat[]{{\includegraphics[width=10cm]{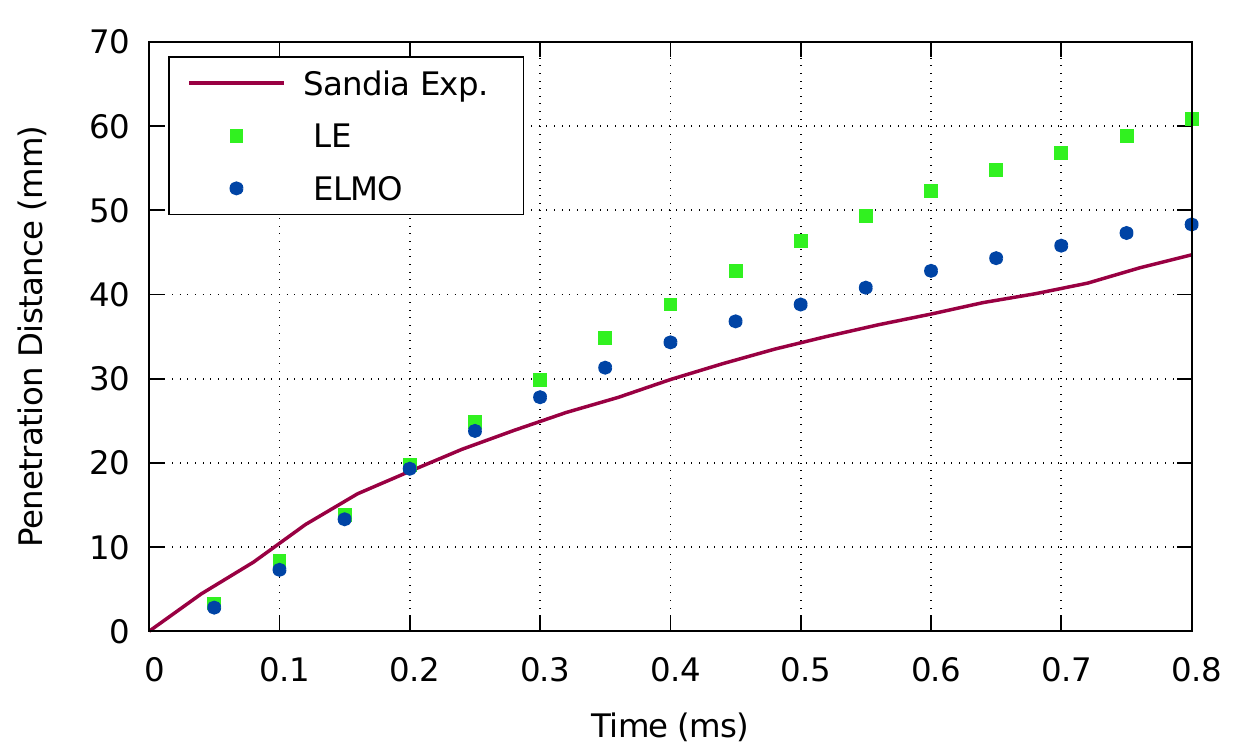} }}%
    \qquad
    \subfloat[]{{\includegraphics[width=10cm]{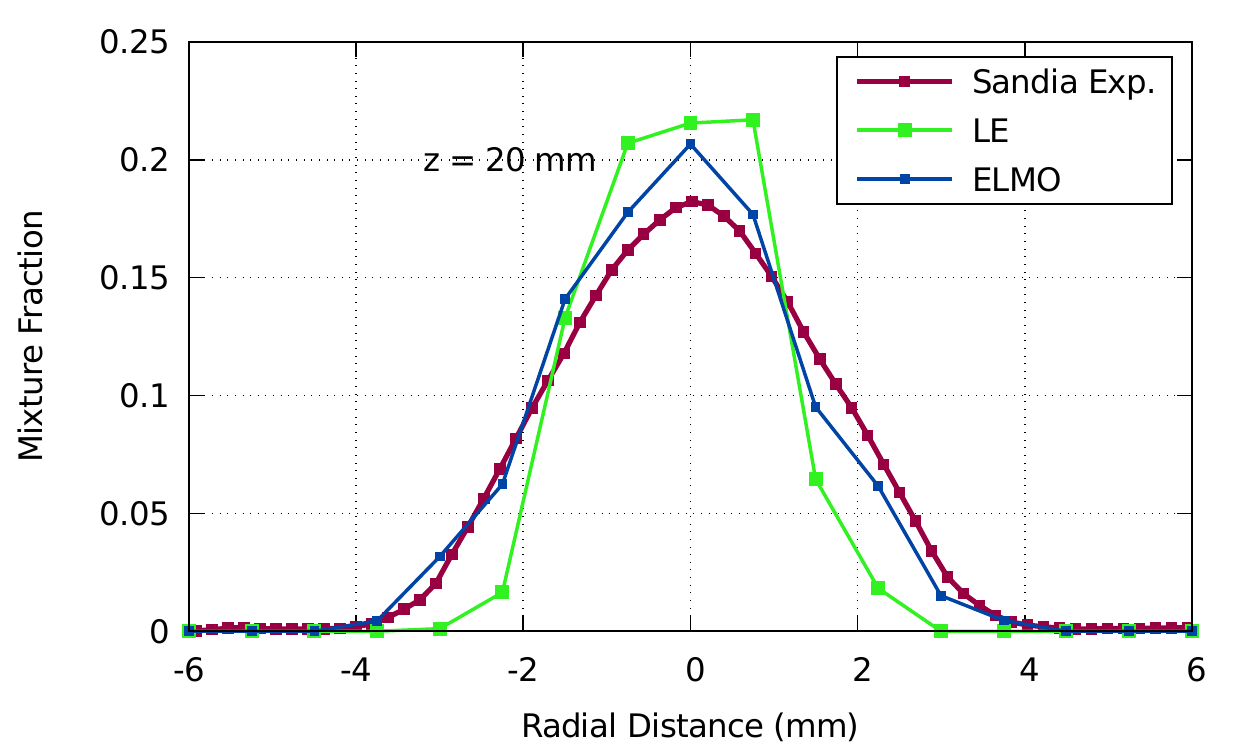} }}%
    \caption{Spray H simulation compared against experimental result of Sandia lab. a) vapor penetration, b) transverse mixture fraction distribution.  Because the experiment has imperfect symmetry, the experimental data for transverse mixing fraction have been shifted horizontally to align with the simulation.}%
    \label{fig:SpH}%
\end{figure}

Figure \ref{fig:SpH} shows (a) the penetration versus time and (b) transverse mass distributions at 1.13 ms after start of injection.  The accuracy of the fuel mixing results is comparable to those of Spray A.  Like Spray A, these experimental data are located far beyond the point where ELMO capsules are terminated, and so the turbulence model also plays a role in these predictions.  Because Spray H is more volatile and the gas temperature so high, the capsules vaporize quickly and the transition to parcels occurs only at very early times, near the location of injection.  For the large majority of the simulation, the fuel vaporizes directly and completely from the capsules. 

%Of note, the errors present in the Spray H results are consistent:  the penetration curve is over-predicted by the ELMO model, indicative of an under-prediction of gas entrainment.  Similarly, the radial dispersion of mixture fraction is less in the simulation than in the experiments, consistent with under prediction of entrainment.  The penetration results, however, were more sensitive to the turbulence model than to the details of the ELMO model.  Whatever the source of the error, it seems to afflict the Lagrangian-Eulerian results in much the same way as the ELMO simulations.

The final test for the ELMO model is the gasoline Spray G condition.  This eight-hole gasoline direct injection uses iso-octane under non-flashing conditions.  This test case is fundamentally different than the two previous, not only because this is a multi-hole injector, but because of the far lower ambient density.  

These gasoline direct injection conditions do not correspond to the diesel-relevant conditions under which Siebers composed the mixing-limited hypothesis.  There are no existing theoretical boundaries that define when the mixing-limited hypothesis is applicable and when it is not, so this test case provides empirical illumination of the model applicability.   In addition, this condition provides a platform to evaluate the suitability of ELMO for multi-hole nozzles where there can be significant interaction between plumes, despite the imposition of conical growth rate at the defined plume angle oriented at the prescribed hole drill angle. 

\begin{table}[h]
 \centering

\begin{tabular}{|l|l| }
\hline
Parameter &Value \\
\hline
Fuel & iso-octane \\
Diameter & 165 [$\mu$m] \\
Duration of Injection & 0.78 [ms] \\
Mass Injected & 1.25 [mg] \\
Ambient Pressure& 0.6 [MPa] \\
Ambient Temperature & 573 [K] \\
Plume Cone Angle & 20 [degrees] \\
Nozzle Angle & 34 [degrees] \\
Fuel Temperature & 363 [K] \\
Coefficient of Area & 0.68 \\
Initial turbulent kinetic energy & $6.4 \cdot 10^{-3} m^2/s^2$ \\
Initial turbulent dissipation & 0.486 $m^2/s^3$\\
\hline
\end{tabular}
\caption{Spray G experimental parameters \cite{moulai2015} \cite{payri2019}}
 \label{table:sprayGexp}

\end{table}

The conditions used in the simulation are reported in Table \ref{table:sprayGexp}.  The two most critical values are the spray angle, taken from the work of Payri et al.\cite{payri2019} and the coefficient of area, from Moulai et al.\cite{moulai2015}.  For spray angle, Payri et al. performed CFD simulations that included the internal flow.  Depending on their choices of model, the range of spray angle was from 20 to 23 degrees.  The plume angle definition used in their work was based on the minimum observable mass fraction.  In the present work, both values of spray angle were simulated with little effect on the results, and the lower value is arbitrarily chosen for generating the figures below. The rate of injection was given as a transient profile as prescribed by the Engine Combustion Network.

\begin{figure}[H]
    \centering
    \includegraphics[width=7.1cm]{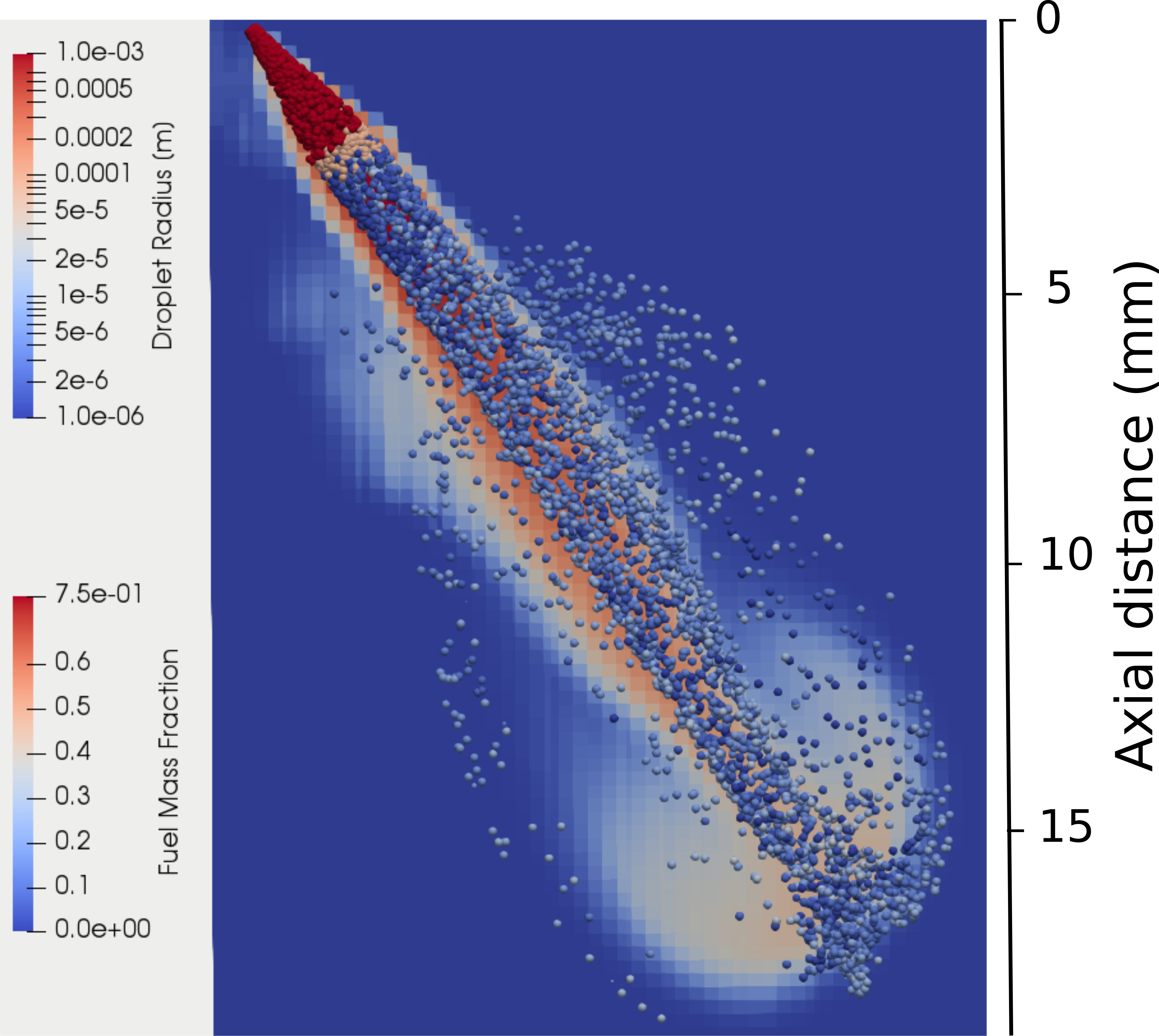}
    \caption{A snapshot of the Spray G fuel injection at 0.35 $ms$ after the start of injection, as modeled by ELMO.  The points with a radius of 1 $mm$ represent points within the capsules whereas the smaller radius points represent parcels.  The filled contours indicate mass fraction of iso-octane.}
    \label{fig:sprayGsnap}
\end{figure}

The eight hole injection was reduced to a one-eighth sector as illustrated in Fig.\ \ref{fig:sprayGsnap}.  This figure also shows that the capsules, indicated by red points, represent a far lower fraction of the liquid fuel lifetime than in the previous diesel cases because less liquid is vaporized at lower ambient temperature and density but the liquid volume fraction decreases by dilution.  Here, the maximum extent of the capsules is about 3 $mm$, at which point the capsules transition to parcels, which further evolve.  

Gasoline direct injection sprays tend to bend inwards, a tendency evident in Fig.\ \ref{fig:sprayGsnap} after the transition to parcels.  The fuel vapor is similarly advected inwards towards the injector axis, indicating entrainment of air towards the centerline of the injector.  The imposed cone angle of the capsules does not prevent the inwards bending experienced by gasoline sprays.  Even though ELMO imposes an initial cone angle, the entrainment that it enforces realistically creates a low pressure zone between plumes. This low pressure zone then deflects the plumes towards the centerline.

\begin{figure}[H]
    \centering
    \subfloat[]{{\includegraphics[width=10cm]{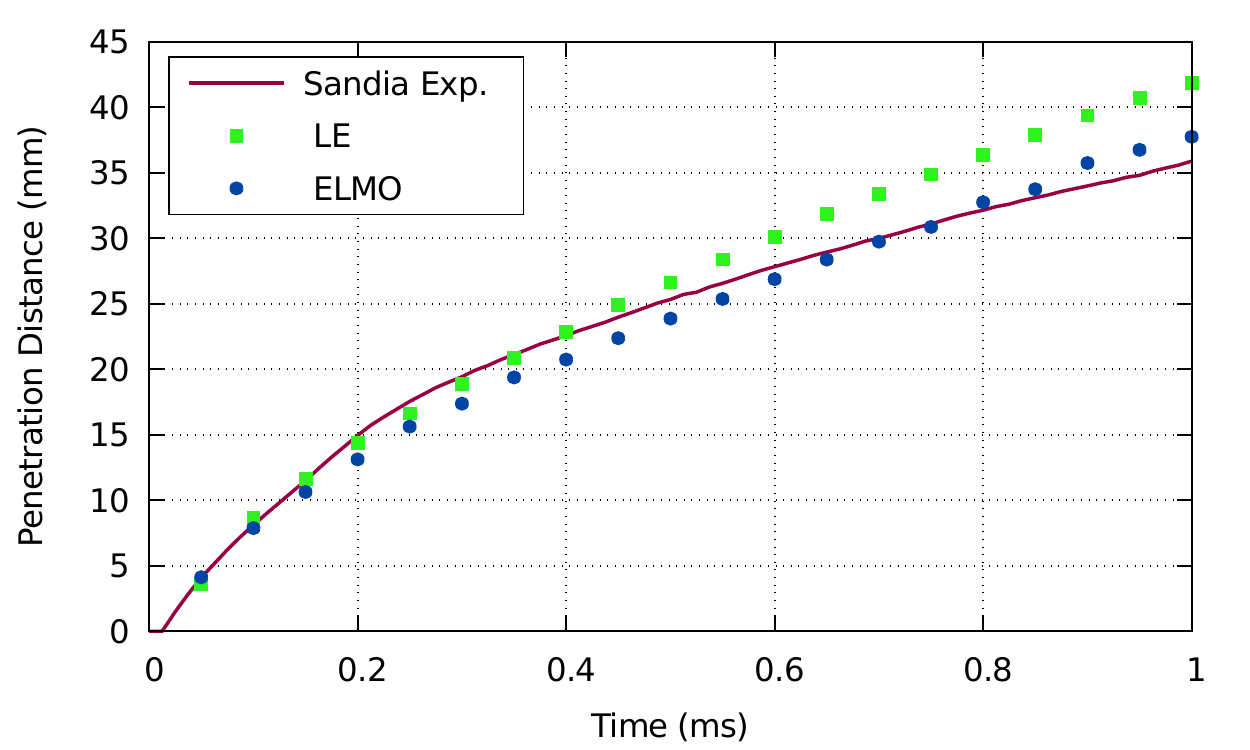} }}%
    \qquad
    \subfloat[]{{\includegraphics[width=10cm]{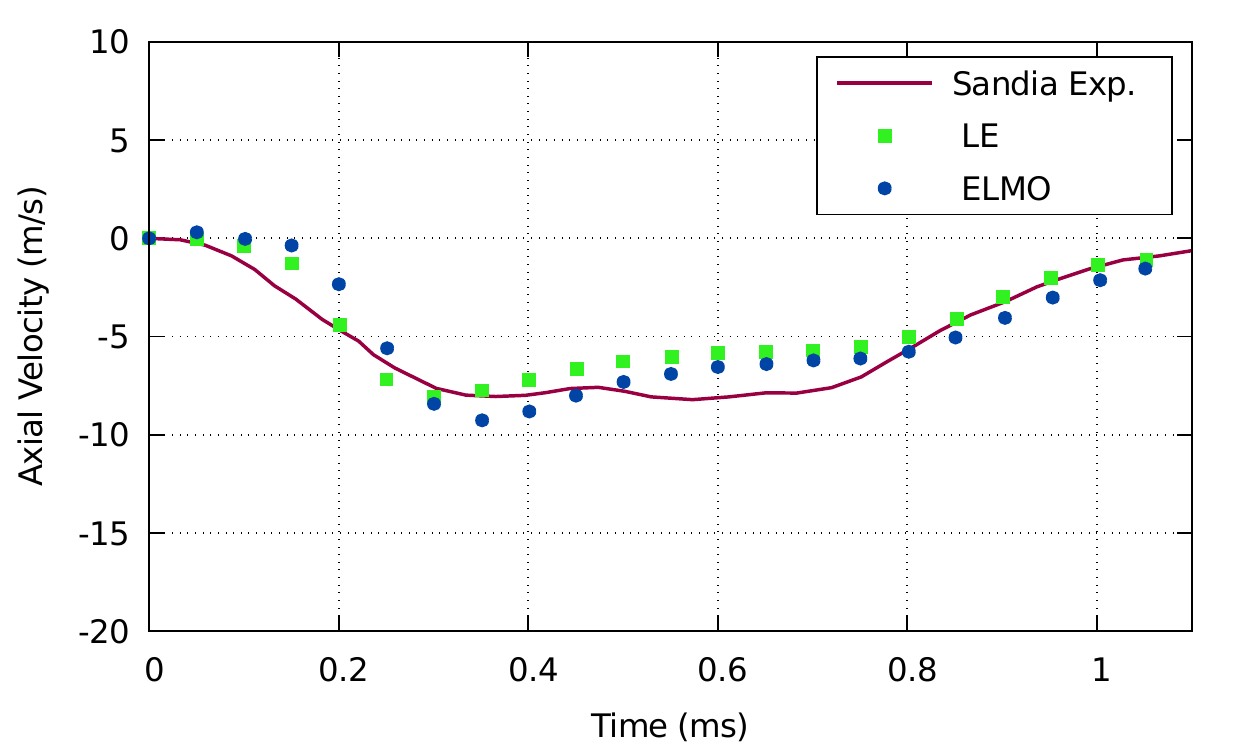} }}%
    \caption{Spray G simulation compared against experimental result of Sandia lab. a) vapor penetration measured in the direction of the injector axis, b) Axial gas velocity at the centerline located at a downstream position of z = 15 $mm$.  Positive values indicate motion away from the injector and negative values indicate motion towards the injector.}%
    \label{fig:SpG}%
\end{figure}

For validation, we use penetration data from the Engine Combustion Network and measured gas velocity data from Sphicas et al. \cite{sphicas2017}.  The penetration data are measured in the direction of the injector axis, not the hole axis, as a function of time.  The gas velocity data comes from transient, non-intrusive measurements made between plumes at a point 15 $mm $ below the injector tip.

Figure \ref{fig:SpG} provides experimental validation for the predicted penetration and the axial gas velocity at a point on the axis of the injector.  The axial velocity validation is based on the observation that the downward spray motion induces temporarily, an upward motion of air along the injector axis.  These velocity data show that shortly after the start of injection, the ambient gas begins moving towards the injector.  The strength of this upward motion is indicative of the momentum exchange between the sprays and the surrounding gas. 

The axial velocity shown in Fig.\ \ref{fig:SpG} has historically been a difficult quantity to predict with Lagrangian spray models.  Sphicas et al. \cite{sphicas2017} simulated Spray G using the same commercial code as is used in the present work (CONVERGE), but with LES turbulence modeling.  As a tunable parameter, they investigated a range of plume cone angles, from 25 to 40 degrees.  They also tried using internal flow simulations to directly map the flow at the exit of the injector orifice to the spray boundary condition.  The strength of the ELMO model is that, without adjustment, it can predict the upward gas flow as shown in Fig.\ \ref{fig:SpG}.  As this flow is driven by interfacial drag, this result is confirmation of the spray model's inertial behavior.

\begin{figure}[H]
    \centering
    \includegraphics[width=0.85\linewidth]{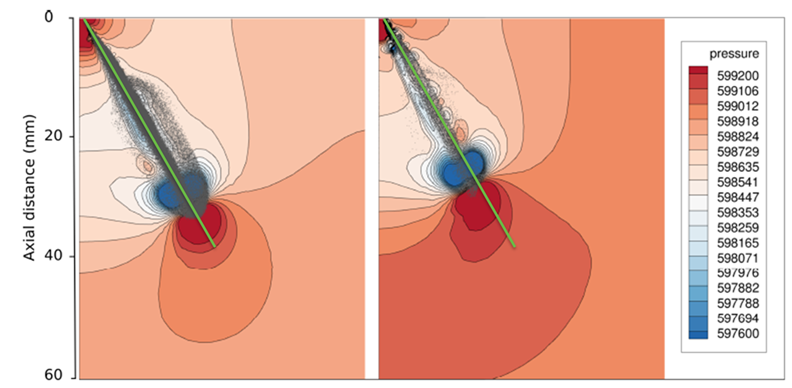}
    \caption{A snapshot of the Spray G pressure field at 0.75 $ms$ after the start of injection, as modeled by Lagrangian Eulerian (left) and ELMO (right). The two lines, placed at the same angle and position, allow comparison of the spray deflection.}
    \label{fig:sprayG_press}
\end{figure}

Because multi-hole injectors have the potential for spray-to-spray interactions, we also chose to look at the pressure field around one of the sprays.  The creation of low pressure regions in the vicinity of the spray plumes is a potential factor in spray bending or spray collapse.  Fig. \ref{fig:sprayG_press} shows the pressure field for both the traditional Lagrangian-Eulerian and ELMO simulations of Spray G.   The green lines, which are intended to make spray deflection more obvious, show that ELMO predicts greater plume deflection than the standard Lagrangian model. Note that this is a significant result because the model construct for an ELMO capsule has fixed inputs for plume direction and growth, and therefore, there is no bend of the capsule itself. However, the Eulerian solution overlaid to the capsule produces expected results. A low pressure zone is established at the injector axis that is more prominent for ELMO.

%The green lines, which are intended to make spray deflection more obvious, show that ELMO predicts greater plume deflection than the standard Lagrangian model.  A low pressure zone is established at the injector axis that is more prominent for ELMO. The pressure field, and probably lower liquid fuel mass, bends the plume more with ELMO.

In these three non-combusting cases, the cost of the CFD calculation was dominated by the gas phase solution.  The cost of both the ELMO model and the traditional Lagrangian-Eulerian spray model were both a small fraction of the overall cost.  In these non-combusting cases, the spray represented roughly eight percent or less of the total computational cost.  When applied to combustion cases in engines, it is anticipated that the computational cost of the spray model would be an even smaller fraction of the total computational time.  
\FloatBarrier
\section*{Conclusions}
A mixing-oriented spray model has been conceived, implemented, and tested.   This new model, ELMO, is based on ideas of thermodynamic and inertial equilibrium in the dense spray core, resulting in an approach that targets mixing-limited conditions. ELMO, as demonstrated by the results generated here, can be applied to diesel sprays and gasoline sprays without adjustment.  The computational cost of the model was roughly equivalent to traditional Lagrangian/Eulerian simulation.  

The ELMO model's ability to predict behavior of gasoline direct injection sprays suggests that, at least within the first several millimeters, gasoline sprays may also be mixing-limited.  This conclusion is only based on the observation of a single gasoline spray case and so must be strongly qualified until further cases can be examined.  Because there is currently no theory that successfully delineates the bounds of the mixing-limited regime, such empirical observations must suffice for the present.

Observed limitations of the model's capabilities indicate areas for future work and development.  The transition from the ELMO capsules to the parcel-based tracking represents an abrupt change in aerodynamic drag.  Consequently, an artificial change in spray angle occurs at that point.  A more graceful transition process would help produce more realistic downstream spray shapes. Analogous improvements for the rate of heat transfer are warranted.

The present work represents an initial foray into making a fully-coupled, mixing-limited, CFD computation.  Unlike prior mixing-limited models in the literature, the ELMO model has more degrees of freedom since the model is two-way coupled.  The comparison to Lagrangian-Eulerian spray modeling is a particular challenge; given the relative maturity of the Lagrangian-Eulerian modeling approach, one expects fairly good experimental agreement from the Lagrangian-Eulerian results.  The fact that the ELMO model produces comparable accuracy in this initial effort is encouraging.

\section*{Acknowledgements}
 Funding for the project was provided by the Spray Combustion Consortium of automotive industry sponsors, including Convergent Science Inc.\, Cummins Inc.\, Ford Motor Co.\, Hino Motors Ltd., Isuzu Motors Ltd., Groupe Renault, and Toyota Motor Co.. Experiments were performed at Combustion Research Facility, Sandia National Laboratory with support from the U.S. DOE Office of Vehicle Technologies. Sandia National Laboratories is a multi-mission laboratory managed and operated by National Technology and Engineering Solutions for Sandia LLC, a wholly owned subsidiary of Honeywell International, Inc.\, for the U.S. Department of Energy's National Nuclear Security Administration under contract DE-NA0003525. The authors also thank
Dr. Yajuvendra Shekhawat of Convergent Science Inc.\ for his help in using the CONVERGE code's full capabilities.

\FloatBarrier

\section*{References Cited}
\bibliographystyle{plain} 
\bibliography{ELMO}

\end{document}